\documentstyle[12pt,epsf]{article}
\def\PR{Phys.\ Rev.\ }

\def\PL{Phys.\ Lett.\ }
\def\NP{Nucl.\ Phys.\ }
\def\IJMP{Int.\ J.~Mod.\ Phys.\ }

\def\rf#1#2#3{{\bf #1}, (#2), #3}
\def\etal{{\it et.\ al.},\ }

\begin{document}
\begin{titlepage}
\rightline{\vbox{\halign{&#\hfil\cr
	 		 & ANL-HEP-CP-00-004\cr
	 		 & \today\cr}}}
\vspace{2.0cm}

\begin{center}
\LARGE
{\bf A Comparison of Spin Observable Predictions for RHIC} \\
\vspace{1.0cm}
\Large
Gordon P. Ramsey
\footnote{To be published in the proceedings of the Circum-Pan-Pacific RIKEN
Workshop on High Energy Spin Physics, RIKEN, Waco, Japan, November 3-6, 1999.
This work is supported in part by the U. S. Department of Energy, Division of
High Energy Physics, Contract W-31-109-ENG-38. E-mail address: gpr@hep.anl.gov}
\\
\medskip
Loyola University Physics Department, Chicago, IL, 60626 USA \\
and \\
Argonne National Lab, Argonne, IL, 60439 USA
\end{center}
\normalsize

\begin{abstract}
There have been many versions of spin-dependent parton distributions in
the literature. Although most agree with present data within uncertainties,
they are based upon different physical assumptions. Some physical models are
discussed and the corresponding predictions for double spin asymmetries are
shown. A summary of the most feasible measurements in the appropriate 
kinematic regions at RHIC, which should yield the most useful information
about the polarized gluon distribution, is given.
\end{abstract}
\end{titlepage}

\section{Introduction}

Recent polarized deep-inelastic experiments have yielded valuable information
about the quark helicity distributions in the proton, but have left some
crucial questions unanswered. One of the key unknowns is the size of
the polarized gluon distribution. In light of the data, there have been
numerous constituent models reflecting contributions of quarks and gluons to
the proton spin.\cite{ggr,gs,bbs,grsv} Most of these models reproduce existing
data fairly well, but vary as to their physical assumptions and the overall
functional form. There is a wide variety of results for the polarized strange
sea and the polarized gluons in the proton, for example. Since the constituent
distributions are used for predicting the spin observables, their measurement
can be used to distinguish between the physical assumptions of these
models. Hence, further experimental information is necessary in order to
understand the nature of these constituents, in particular, the polarized glue.

The Relativistic Heavy Ion Collider (RHIC) at Brookhaven (BNL) is well suited
for the polarized beam experiments which can help to reveal the nature of
the proton constituents' spin properties. The two major detectors, STAR and
PHENIX, cover a similar kinematic region, but STAR has wider angular coverage,
while PHENIX has finer granularity for more precise measurements. 

This paper discusses the optimal observables for these detectors to determine
$\Delta G$. Section II will provide a brief overview
of some theoretical models of the spin constituents of the proton. The physical
assumptions and use of present polarized DIS data will be discussed. In
section III, key experimental predictions for the determination of $\Delta G$
will be compared. Finally, suggestions are made for the best possibilities to
experimentally determine the nature of $\Delta G$ in the accessible kinematic
regions of the two detectors.

\section{Models of the Polarized Distributions}

In recent work\cite{ggr,gr} we constructed three sets of polarized parton
distributions for the valence and sea quarks, based upon three models for the
polarized gluons. Each gluon model has a different physical basis, but 
the quark distributions comform to a set of reasonable theoretical assumptions.
All of the distributions are generated at $Q_0^2=1.0$ GeV$^2$ and evolved in
NLO, entirely in $x$-space, up to the $Q^2$ values necessary to predict the
spin observables. All three sets are in good agreement with data.

The polarized valence distributions are generated from a modified SU(6)
distribution with a spin dilution factor to control their small-$x$ behavior.
The polarized valence distributions are written in terms of the unpolarized 
CTEQ distributions as:
\begin{eqnarray}
\Delta u_v&=& \cos \theta_D \bigl[u_v-\frac{2}{3}d_v\bigr] \nonumber \\
\Delta d_v&=& \cos \theta_D \bigl[-\frac{1}{3}d_v\bigr],
\end{eqnarray}
where the spin dilution factor is: $\cos \theta_D=[1+R_0(1-x)^2/\sqrt(x)]^{-1}$.
The free parameter, $R_0$, is fixed by applying the Bjorken Sum Rule. In the 
$Q^2$ region of the present PDIS data, we find that $R_0\approx \frac{2}{3}
\alpha_s$. This parametrization gives the appropriate behavior at both small
and large $x$.

For the polarized sea, we assume a broken SU(3) model, to account for mass
effects in polarizing the strange sea. Our models separate all light flavors
in the valence and sea. Charm is included via the evolution equations ($N_f$),
at the appropriate $Q^2$ of charm production, to avoid any non-empirical 
assumptions about its size. The small-$x$ behavior is of the Regge type, and
the large-$x$ behavior is compatible with the appropriate counting 
rules.\cite{bbs}

The polarized sea distributions are extracted from both the unpolarized CTEQ
sea  distributions and polarized deep-inelastic-scattering data.\cite{gr}
We assume a model of the sea which obtains its polarization from gluon 
Bremsstrahlung, so the net polarization of the quarks is dependent upon their
density in hadrons in LO. We therefore assume that these densities are
directly proportional and the flavor dependent sea distributions have the form
\begin{equation}
\Delta q_i=\eta_i(x) \cdot x \cdot q_i(x).
\end{equation}
The $\eta_i$ are determined from the integrated distributions and sum rules
used to analyze polarized deep-inelastic-scattering data.
We have chosen the functional form of $\eta_i(x)$ to have a reasonable Regge
type behavior, to be consistent with positivity constraints and to yield the
proper normalization (indicated here by the factor $\eta$) in reflecting the
relative spin that each flavor contributes to that of the proton. Here,
$\eta_i(x)$ may be interpreted as a modification of $\Delta q(x)$ due to
unknown soft effects at small-$x$. The normalization requires that 
$\int_0^1 \Delta q_i(x)\cdot dx=\int_0^1 \eta_i(x) \cdot xq_i(x) \cdot dx=\eta
\cdot \int_0^1 xq_i(x)\cdot dx.$
The overall parametrization for each of the polarized sea flavors, including
the $\eta(x)$ functions, the anomaly terms and the up-down unpolarized
asymmetry term can be written (with the CTEQ basis) in the form:
\begin{equation}
\Delta q_i(x)=-Ax^{-0.143}(1-x)^{8.041}(1-B\sqrt{x})\Bigl[1+6.112x+P(x)
\Bigr].
\end{equation}
The values for the variables in this equation are given for each flavor and
each gluon model in Table I. Note that $\delta(x)\equiv 0.278x^{0.644}$ is
due to the asymmetry of the unpolarized up and down anti-quarks.

\begin{table} \caption{Parametrizations for Polarized Sea Flavors}
\begin{tabular}{|c||c||c|c|c|}
\hline
  Flavor    &$\Delta G$  	&A  	&B  	&P(x) 			\cr
\hline                                                  
\hline
  $\Delta \bar u$ &xG 	&0.317 &1.124 &$-\delta(x)-1.682x^{0.937}
(1-x)^{-3.368}(1+4.269x^{1.508})$	\cr
\hline
  $\Delta \bar d$ &xG 	&0.317 &1.124 &$+\delta(x)-1.682x^{0.937}
(1-x)^{-3.368}(1+4.269x^{1.508})$	\cr
\hline
  $\Delta s$ 	&xG 	&0.107 &1.257 &$-3.351x^{0.937}(1-x)^{-3.368}
(1+4.269x^{1.508})$	\cr
\hline
  $\Delta \bar u$ &0 	&0.386 &0.990 &$-\delta(x)$	\cr
\hline
  $\Delta \bar d$ &0 	&0.386 &0.990 &$+\delta(x)$	\cr
\hline
  $\Delta s$ 		&0 	&0.173 &1.070 &0			\cr
\hline
  $\Delta \bar u$ &$<0$ &0.414 &0.954 &$-\delta(x)-10.49x^{1.143}
(1-x)^{-1.041}(1+0.474\ln{x})$	\cr
\hline
  $\Delta \bar d$ &$<0$ &0.414 &0.954 &$+\delta(x)-10.49x^{1.143}
(1-x)^{-1.041}(1+0.474\ln{x})$	\cr
\hline
  $\Delta s$ 	&$<0$ 	&0.212 &0.997 &$-20.89x^{1.143}(1-x)^{-1.041}
(1+0.474\ln{x})$			\cr
\hline
\end{tabular}
\end{table} 

We consider three distinct models for the polarized gluons, which have a 
moderate size. There exists no empirical evidence that the polarized gluon
distribution is large at the relatively small $Q^2$ values of the data. Data
from the Fermilab E704 experiment indicate that it is likely small at these
$Q^2$ values. In addition, a theoretical model of $\Delta G$ based on counting
rules, implies that $\Delta G\approx \frac{1}{2}$.\cite{bbs} Our choice of 
models effectively includes two separate factorization schemes, Gauge 
Invariant (GI or $\overline{MS}$) and Chiral Invariant (CI or Adler-Bardeen),
which can be used to represent the polarized sea distributions.\cite{cheng}

The first set of $\Delta q_i(x)$ functions, quoted in Table I assumes a 
moderately polarized glue, normalized to $\frac{1}{2}$ using the CTEQ
unpolarized gluons. The second polarized gluon model assumes $\Delta G=0$.
This is  equivalent to writing the quark distributions in the gauge-invariant
scheme, since the anomaly term vanishes. The third model is motivated by an 
instanton-induced polarized gluon distribution, which gives a negatively 
polarized glue at small-$x$.\cite{koch} The three polarized gluon 
distributions are written as
\begin{eqnarray}
\Delta G(x)&=&xG(x) \nonumber \\
\Delta G(x)&=&0 \\
\Delta G(x)&=&7 (1-x)^7\bigl[1+0.474\ln(x)\bigr]. \nonumber 
\end{eqnarray}
Our overall distributions agree very well with existing data.

The models of Gehrmann and Stirling are based upon an SU(3) symmetric sea
with the small-$x$ behavior of the sea and glue assumed equal. All of the
difference between the Ellis-Jaffe sum rule and the data are attributed
to $\Delta G$, with $\Delta s\to 0$. This differs considerably with our models.
They also have three different polarized gluon models, which run the range of
hard and soft gluons. The GSA $\Delta G$ is much larger than that of our first
model. The other two fall within the range of ours, and are therefore not 
discussed here. Our motive is to present a wide range of models for $\Delta G$
so that the experiments can be used to distinguish the size of $\Delta G$ as
opposed to the overall accuracy of a particular model. The differences in the
predictions of the GSA and the three GGR models (A, B, and C)\cite{ggr}
are thus due to the $x$-behavior of the quark distributions and the relative
sizes of $\Delta G$.

\section{A Comparison of Experimental Predictions}

The polarization experiments planned for RHIC show great potential for
extracting information on polarized distributions, especially $\Delta G$.
With polarized beams of $70\%$ polarization and luminosity of $2\times 10^{32}$
/cm$^{-2}$/sec$^{-1}$, both prompt-$\gamma$ production and jet production can
be done in a kinematic region where determination of $\Delta G$ is possible.
If the planned integrated luminosity of 320 pb$^{-1}$ at $\sqrt s=200$ GeV is
attained, the resulting data should be good enough to distinguish among many
of the polarized gluon models which have been proposed.

The STAR detector will have a wide angular range to cover a large rapidity,
especially for jet production. The PHENIX detector has a narrow rapidity, but
finer granularity, and is well suited for measuring high $p_T$ prompt photons.
Both are planned to have an accessible $p_T$ range of $10\le p_T\le 30$ GeV
at $\sqrt s=200$ GeV. The predictions shown here cover this kinematic range.
Possible experiments which would provide a measure of $\Delta G$ include:
\begin{itemize}
\item one and two jet production in $e-p$ and $p-p$ collisions,
\cite{gs,goram}
\item prompt photon production\cite{gs,goram}
\item charm production\cite{bbg}
	\begin{itemize}
	\item polarized heavy quark production\cite{bg}: these asymmetries
	are between one and four percent for $\sqrt s=200$ GeV with $p_T\ge 5$
	GeV
	\item $\chi_{0c}$ and $\chi_{1c}$ production\cite{morii}: for very
	large models of $\Delta G$, these asymmetries range from eight down to
	one percent for $\sqrt s=500$ GeV and $2\le p_T\le 30$ GeV.
	\item $\chi_{2c}$ production\cite{gs}: even for the larger models
	of $\Delta G$, these asymmetries range from zero to four percent at
	$\sqrt s=200$ GeV and $2\le p_T\le 8$ GeV.
	\end{itemize}
\item $J/\psi$ production \cite{tt} these asymmetries are typically from 
	two to six percent at $\sqrt s=200$ GeV and $2\le p_T\le 10$ GeV.
\item pion production \cite{rs}.
\end{itemize}

All but the first two of these yield small asymmetries, even for moderately
sizable gluon polarizations. Thus, we feel that jet production and 
prompt-$\gamma$ production are the best choices for extracting information about
$\Delta G$. The asymmetries are not large everywhere, but there are kinematic
regions where models based upon the different sizes of $\Delta G$ can be
distinguished. We have calculated prompt-$\gamma$ production in LO and NLO,
and jet production at LO. The figures compare the predictions of our three
models with the Gehrmann and Stirling A model.

Figure 1 compares the predictions of the GSA (large $\Delta G$) with the
three GGR gluon models for the prompt photon asymmetry, $A_{LL}^{\gamma}$.
The error bars shown are those expected for $A_{LL}^{\gamma}$ at the PHENIX
detector in these kinematic regions. Figure 2 shows the same asymmetry
predictions at fixed $p_T$ of $15$ GeV as a function of rapidity. The error
bar shown is for PHENIX, which operates at essentially this point of rapidity.
STAR has a much wider rapidity coverage, $-1\le y\le 2$. Figure 3 shows the
comparison of the four models for jet production at $\sqrt S=200$ GeV for
zero rapidity. The asymmetries for jet production at $\sqrt S=500$ GeV are
much smaller than for the lower energy.

\begin{figure}
\unitlength1cm
\begin{minipage}[t]{6.5cm}
\begin{picture}(6.0,6.5)
	\hbox{\epsfxsize=6.0cm\epsffile{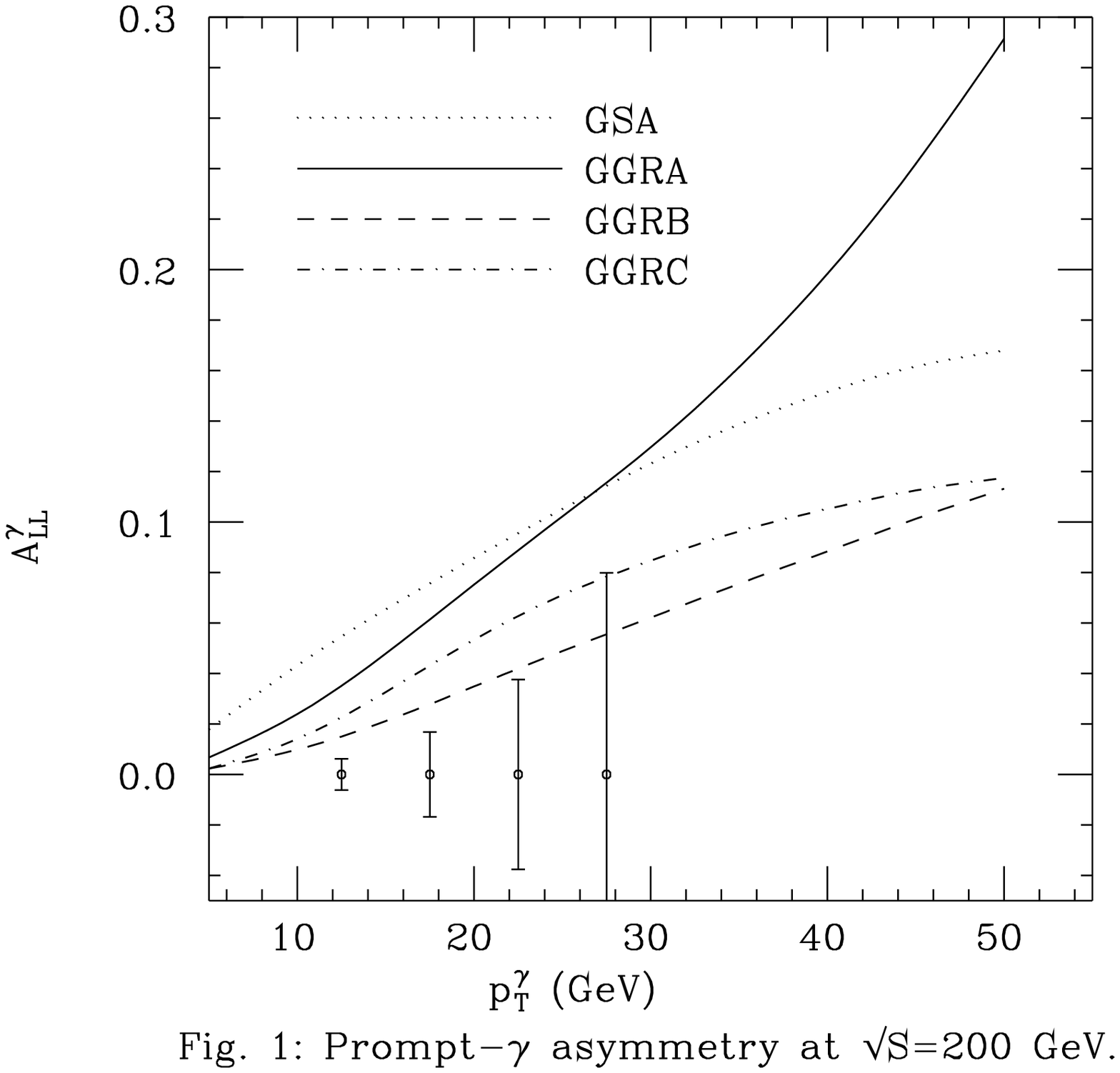}}
\end{picture}
\end{minipage}
\hfill
\begin{minipage}[t]{6.5cm}
\begin{picture}(6.0,6.5)
	\hbox{\epsfxsize=6.0cm\epsffile{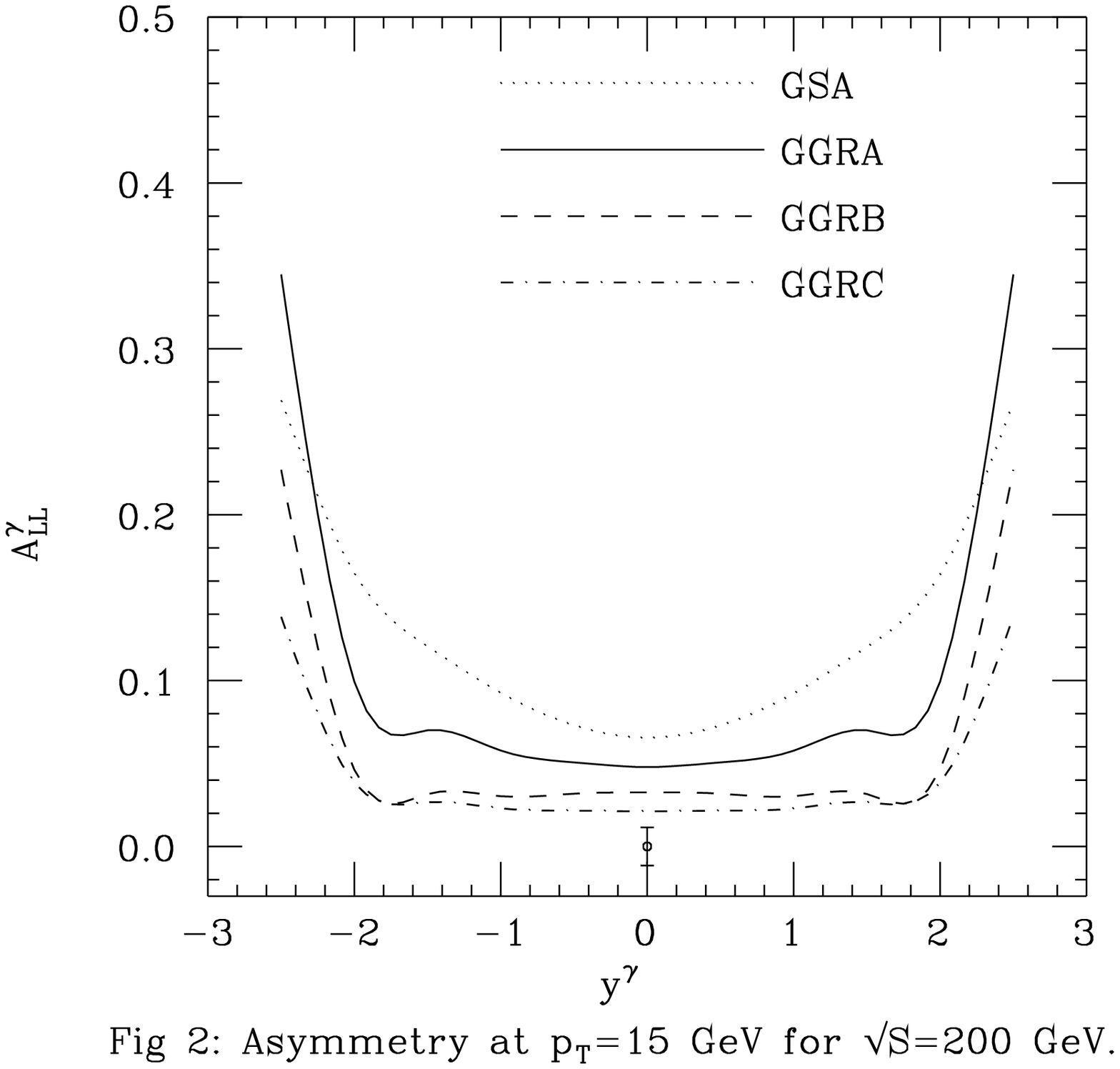}}
\end{picture}
\end{minipage}
\end{figure}

\begin{figure}
 {\hskip 2.0cm}\hbox{\epsfxsize=7.5cm\epsffile{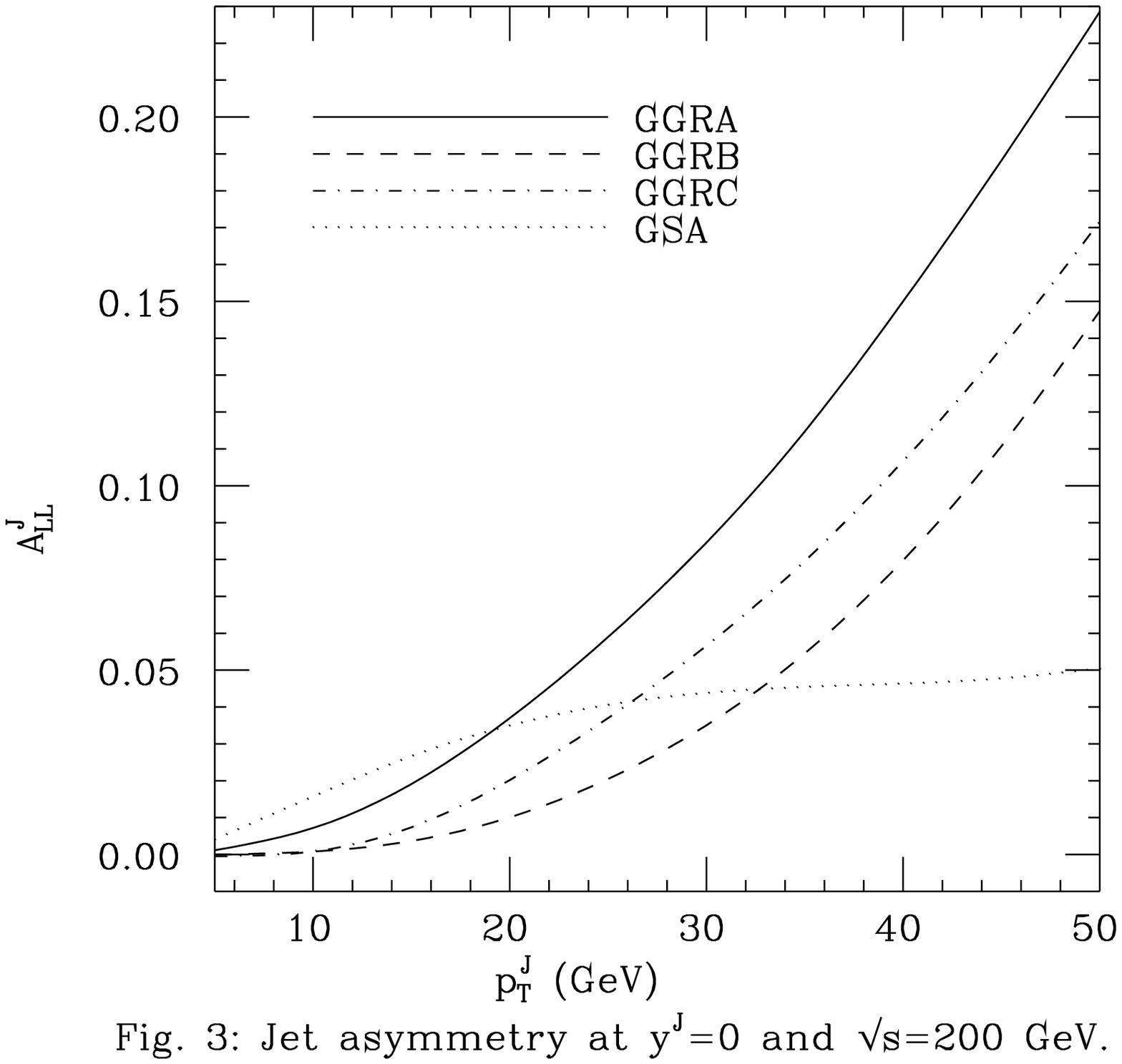}}
\end{figure}

\section{Extracting $\Delta G$}

The various polarized gluon models have different physical bases and provide a
reasonable range of possibilities, which can be narrowed down by future 
experiments at RHIC. If the polarized gluon distribution is moderately 
positive at $Q^2=1$ GeV$^2$, the asymmetry for prompt photon production is
among the best candidates for determining the size of $\Delta G$. Jet 
production will be a close contender for distinguishing among the various
predictions for $\Delta G$. Much depends upon the relative errors in the 
applicable kinematic regions of STAR and PHENIX.

According to the projected uncertainties for STAR and PHENIX, the most 
favorable region to study prompt photon production is for $15\le p_T\le 25$
GeV at $\sqrt s=200$ GeV. Although the asymmetries are closer together here,
the favorable small uncertainties should be able to separate the large and
small models for $\Delta G$ (Fig. 1). Also, at $p_T=15$ GeV, the large 
rapidity region $1\le \mid y\mid \le 2$ is a favorable place for STAR to 
measure the asymmetry due to the separation of predictions in the models (Fig.
2). Since PHENIX is designed for the small rapidity region, the measurement of
$A^{\gamma}_{LL}$ with the better uncertainties is also promising, and will
provide a good cross check of the results obtained by STAR. 

Jet production is also a good candidate for determination of whether $\Delta G$
is large or small. Since the asymmetries are fairly close together in the
$p_T$ region between $15$ and $30$ GeV, measurements of this asymmetry require
the larger values of $p_T$ to distinguish the relative size of $\Delta G$
extracted from these predictions (Fig. 3).

Prompt photon and jet production, measured by STAR and PHENIX in these 
kinematic regions, appear to be the best candidates for providing an 
indication of the nature of $\Delta G$.

\vskip3mm
\section*{Acknowledgements}
These results are based upon work done with L.E. Gordon (Jefferson Lab) and
M. Goshtasbpour (Shahid Beheshti Univ, Tehran).

\end{document}